\documentstyle[12pt,epsf]{article}
\begin{document}
\renewcommand{\thetable}{\Roman{table}}
\def \beq{\begin{equation}}
\def \eeq{\end{equation}}
\def \G{{\rm GeV}}
\def \gth{g_{\theta}}
\def \M{{\rm MeV}}
\def \shat{\hat s}
\rightline{FERMILAB-PUB-95/394-T}
\rightline{EFI-95-75}
\rightline{hep-ph/9512299}
\vspace{0.5in}
\centerline{\bf FORWARD-BACKWARD ASYMMETRIES}
\centerline{\bf IN HADRONICALLY PRODUCED LEPTON PAIRS
\footnote{To be submitted to Phys.~Rev.~D.}}
\vspace{0.5in}
\centerline{\it Jonathan L. Rosner}
\centerline{\it Theoretical Physics Division}
\centerline{\it Fermi National Accelerator Laboratory, Batavia, IL 60510}
\bigskip
\centerline{and}
\bigskip
\centerline{\it Enrico Fermi Institute and Department of Physics}
\centerline{\it University of Chicago, Chicago, IL 60637
\footnote{Permanent address.}}
\bigskip

\centerline{\bf ABSTRACT}
\medskip
\begin{quote}
It has now become possible to observe appreciable numbers of hadronically
produced lepton pairs in mass ranges where the contributions of the photon and
$Z^0$ are comparable.  Consequently, in the reaction $p \bar p \to \ell^-
\ell^+ + \ldots$, substantial forward-backward asymmetries can be seen.  These
asymmetries provide a test of the electroweak theory in a new regime of
energies, and can serve as diagnostics for any new neutral vector bosons
coupling both to quarks and to charged lepton pairs.
\end{quote}
\bigskip

\centerline{\bf I.  INTRODUCTION}
\bigskip

The reaction $p \bar p \to \ell^- \ell^+ + \ldots$ is dominated by virtual
photons at low energy \cite{DY}, by the $Z^0$ at $m(\ell^- \ell^+) = M_Z$, and
by photon-$Z$ interference everywhere else.  Here $\ell = (e, \mu)$ stands for
an isolated charged lepton, i.e., one not due to charm or bottom semileptonic
decay.  For lepton pair masses between about 60 and 80 GeV/$c^2$, the similar
magnitude of photon and $Z$ contributions leads one to expect an asymmetry of
about $-50\%$, with three out of four $\ell^+$ having greater rapidity than
$\ell^-$ in the direction of the incident proton.  For pair masses above 100
GeV/$c^2$, one expects an asymmetry of about $+ 50\%$, with three out of four
$\ell^-$ having greater rapidity than $\ell^+$ in the direction of the proton.

Several years ago, when collider experiments at CERN (c.m. energy 540 GeV) and
Fermilab (c.m. energy 1.8 TeV) each had accumulated several events per picobarn
of cross section, the observation of these asymmetries was estimated to be
possible at the $2 \sigma$ level \cite{asy}.  Now, with samples from the CDF
and D0 detectors at Fermilab of order 100 pb$^{-1}$, it should be possible to
observe these interference effects unambiguously.  This would be the first
study of such effects at masses above the $Z$, a range which has only recently
become available to the LEP $e^+ e^-$ collider.

In Ref.~\cite{asy} it was mentioned that new neutral gauge bosons beyond the
photon and $Z$ would lead to deviations of the forward-backward asymmetry from
that predicted in the standard model \cite{LRR,offpk,BDRW}.  Compositeness
(i.e., a new effective four-fermion interaction) also can affect the asymmetry
\cite{RS}.  The present note investigates these points with regard to the reach
of the current Fermilab experiments.  The effects associated with new gauge
bosons $Z'$ are found to be modest until one approaches pair masses $m(\ell^-
\ell^+) \simeq M_{Z'}$, where the asymmetries can provide distinction among a
number of possibilities.  The compositeness signatures in forward-backward
asymmetries turn out to be of utility comparable to (but not exceeding) direct
searches for excess production cross sections at high lepton pair masses.

In Sec.~II we tour of the zoo of extra $Z$'s, noting present limits from direct
searches.  We then choose in Sec.~III a $Z'$ whose mass is compatible with
these limits, and ask for its effects on the forward-backward asymmetry. A
brief discussion of the achievable limits on compositeness occupies Sec.~IV,
while Sec.~V concludes.
\bigskip

\centerline{\bf II.  A BRIEF TOUR OF EXTRA $Z$'s}
\bigskip

\leftline{\bf A.  Unifying groups}
\bigskip

We use the notation of Refs.~\cite{LRR}, \cite{offpk}, and \cite{RR}, which may
be consulted for further details.  Other discussions of extra $Z$'s may be
found in Refs.~\cite{LR} and \cite{other}.

The standard SU(3) $\times$ SU(2) $\times$ U(1)$_Y$ model may be incorporated
into an SU(5) \cite{GG}, with the known quarks and leptons in each family
belonging to the representations {\bf 5}$^*$ or {\bf 10} of the unifying group.
 These may be combined into a single 16-dimensional representation of SO(10)
\cite{SO}, with the addition of a right-handed neutrino.  The group SO(10)
contains SU(5) $\times$ U(1) as a subgroup; we shall denote this U(1) by the
subscript $\chi$, and its corresponding gauge boson by $Z_\chi$.

A further embedding into E$_6$ is suggested by some string-theory models
\cite{ST}; the U(1) which arises when E$_6$ breaks down to SO(10) $\times$ U(1)
will be denoted by the subscript $\psi$, and its corresponding boson by
$Z_\psi$.  The 15 known fermions in each family of the standard model belong to
27-plets of E$_6$, consisting of {\bf 16}, {\bf 10}, and {\bf 1}
representations of SO(10). The {\bf 16}, as mentioned, contains the standard
fermions and a right-handed neutrino.  The {\bf 10} contains weak isosinglet
quarks and antiquarks of charge $\pm 1/3$, and weak doublets of leptons and
antileptons.  The {\bf 1} contains an isosinglet Majorana neutrino.

The most general $Z'$ within E$_6$ then may be parametrized \cite {LR} as
\beq \label{eqn:th}
Z' = Z_\psi \cos \theta + Z_\chi \sin \theta~~~.
\eeq
When $\theta = {\rm arctan}(\sqrt{3/5}) = 37.78^{\circ}$, the corresponding
$Z'$ is denoted by $Z_\eta$, and corresponds to a specific breaking of E$_6$
suggested by some superstring theories \cite{ST}.  When $\theta = {\rm
arctan}(-\sqrt{5/3}) = 127.78^{\circ}$, the corresponding $Z'$ is the one which
arises when E$_6$ breaks down to SU(6) $\times$ SU(2)$_I$.  The subscript
stands for ``inert,'' since all gauge bosons of SU(2)$_I$ are neutral.  The
$I_{3I} = 0$ member of the SU(2)$_I$ triplet is called $Z_I$.

We shall assume in what follows that a single extra U(1) beyond the standard
model remains unbroken at energies accessible to present accelerators. Slightly
different patterns arise from other symmetry breaking schemes, such as SO(10)
$\to$ SU(4) $\times$ SU(2)$_L \times$ SU(2)$_R \to $ SU(3)$_{\rm color} \times$
U(1)$_{B-L} \times$ SU(2)$_L \times$ U(1)$_R$, if the SU(4) and SU(2)$_R$ break
at different scales.  We shall also assume that the mixing between the ordinary
$Z$ and the $Z'$ remains extremely small.  Stringent constraints on this mixing
exist, as discussed in some of the more recent Refs.~\cite{other}.
\bigskip

\leftline{\bf B.  Couplings}
\bigskip

The vector and axial-vector couplings $C_V$ and $C_A$ of the photon, $Z$, and
$Z'$ to $u$ and $d$ quarks and to charged leptons have been given in
Ref.~\cite{offpk}.  Equivalently, the left-handed and right-handed couplings
$C_L \equiv C_V - C_A$ and $C_R \equiv C_V + C_A$ are shown in Table I.  Here
we denote
$$
g_Z^2 \equiv e^2/[x(1-x)]~~,~~x \equiv \sin^2 \theta_W \simeq 0.231~~~,
$$
\beq \label{eqn:coup}
\gth^2 \equiv \frac{5}{3}e^2/(1-x)~~~,
\eeq
\beq
A \equiv \cos \theta/2 \sqrt{6}~~,~~~B \equiv \sin \theta/2 \sqrt{10}~~~.
\eeq
We have assumed a universal U(1) coupling strength \cite{LR} which would be
that arising if E$_6$ broke in a single step to the standard model $\times$
U(1)$_\theta$.

\begin{table}
\begin{center}
\caption{Left- and right-handed couplings.}
\begin{tabular}{l c c c c c c} \hline \hline
 & \multicolumn{6}{c}{Fermion} \\
 & \multicolumn{2}{c}{$u$ quark} & \multicolumn{2}{c}{$d$ quark} &
\multicolumn{2}{c}{Electron} \\
Boson & $C_L$ & $C_R$ & $C_L$ & $C_R$ & $C_L$ & $C_R$ \\ \hline
$\gamma$ & $2e/3$ & $2e/3$ & $-e/3$ & $-e/3$ & $-e$ & $-e$ \\
$Z$ & $g_Z(-\frac{1}{2} + \frac{2}{3}x)$ & $g_Z( \frac{2}{3}x)$ &
      $g_Z( \frac{1}{2} - \frac{1}{3}x)$ & $g_Z(-\frac{1}{3}x)$ &
      $g_Z( \frac{1}{2} -     x       )$ & $g_Z    (-x)       $ \\
$Z'$ & $-\gth(A+B)$ & $\gth(A+B) $ &
              $-\gth(A+B)$ & $\gth(A-3B)$ &
              $\gth(3B-A)$ & $\gth(A+B) $ \\ \hline \hline
\end{tabular}
\end{center}
\end{table}
\bigskip

\leftline{\bf C.  Decay width}
\bigskip

If all members of three 27-plets of E$_6$ can be produced in the decay of a
heavy $Z'$, there is a simple expression for its width \cite{BDRW}:
\beq
\Gamma(Z') = \frac{5}{2} \alpha (M_{Z'}) M_{Z'} /(1-x)~~~,
\eeq
where $\alpha (M_{Z'})$ is the electromagnetic fine-structure constant
evaluated at $M_{Z'}$.  This corresponds to a ratio $\Gamma/M \simeq 2.5\%$, so
the $Z'$ should be relatively narrow.  Its precise width will depend on the
availability of open channels.  Some members of the 27-plet could be too heavy
to be produced in pairs in the decays of $Z'$, reducing its predicted width.
Alternatively, the presence of superpartners could increase the predicted
width.
\bigskip

\leftline{\bf D.  Scattering amplitudes}
\bigskip

The vector- and axial-vector nature of the interaction with gauge bosons
implies that the annihilation process $f \bar f \to e^- e^+$ may be uniquely
specified by the helicities of the initial fermion $f$ and final electron
$e^-$.  We may then express the corresponding amplitudes, generalizing the
result of Ref.~\cite{asy}, as
$$
A_{ij} \equiv A(f_i \bar f \to e^-_j e^+) = -Q e^2 + \frac{\shat}
{\shat - m_Z^2 + i M_Z \Gamma_Z} C_i^Z(f) C_j^Z(e)
$$
\beq \label{eqn:amp}
+ \frac{\shat}{\shat - m_{Z'}^2 + i M_{Z'} \Gamma_{Z'}} C_i^{Z'}(f)
C_j^{Z'}(e)~~~.
\eeq
Here $\shat$ denotes the square of the c.m. energy, while the coefficients are
those given in Table I for $(i,j) = (L,R)$. The differential cross section for
$f \bar f \to e^- e^+$ is then
$$
\frac{d \sigma(f \bar f \to e^- e^+)}{d \cos \theta^*} = (1/128 \pi \shat)
[ (|A_{LL}|^2 + |A_{RR}|^2)(1 + \cos \theta^*)^2
$$
\beq \label{eqn:diff}
+ (|A_{LR}|^2 + |A_{RL}|^2)(1 - \cos \theta^*)^2]~~~.
\eeq
The forward and backward cross sections $\sigma_F$ and $\sigma_B$ are those
obtained by integrating (\ref{eqn:diff}) over positive and negative values of
$\cos \theta^*$.  The forward-backward asymmetry is then
\beq
A_{FB} \equiv \frac{\sigma_F - \sigma_B}{\sigma_F + \sigma_B}
= \frac{3}{4}\frac{|A_{LL}|^2 + |A_{RR}|^2 - |A_{LR}|^2 - |A_{RL}|^2}
{|A_{LL}|^2 + |A_{RR}|^2 + |A_{LR}|^2 + |A_{RL}|^2}~~~.
\eeq
\newpage

\leftline{\bf E.  Present experimental limits and available data}
\bigskip

The CDF Collaboration \cite{CDFZlims} has published lower limits on various
types of $Z'$ bosons including the ones of the type considered here.  The 95\%
c.l. lower limits on $(Z\psi,~Z_\eta,~Z_\chi,~Z_I)$ masses are (415, 440, 425,
400) GeV/$c^2$, respectively.  In this data set, based on 19.7 pb$^{-1}$, there
are 40 $e^+ e^-$ events above an invariant mass of 125 GeV/$c^2$.  The present
data sample is approximately 5 times as large, and one can make use of dimuon
as well as dielectron data. All told, one can expect at least a factor of ten
statistical improvement with respect to the results of Ref.~\cite{CDFZlims}
once the current run concludes, or an improvement with respect to the results
noted in Ref.~\cite{asy} by a factor of about 40. What can one learn from such
a data sample?

First of all, the asymmetry below the $Z$ (in the 60 - 80 GeV/$c^2$ range)
should be very pronounced.  (See especially Figs.~3(c) and 4(b) of
Ref.~\cite{asy}.)  However, it is necessary to take account of radiative
corrections in order to compare data with predictions.  The asymmetry at the
$Z$ itself is small \cite{CDFZasy} and rapidly varying with lepton pair mass.
If an electron or positron initially in the $Z$ peak loses energy through
undetected radiation, it can appear to belong to a lower-mass pair.

The observation of an asymmetry at lepton pair masses above the $Z$ is much
less dependent on radiative corrections.  Here, as we shall see in Sec.~III and
has been shown in Ref.~\cite{asy}, the asymmetry is expected to be large and
slowly varying with pair mass.  With a few hundred lepton pairs anticipated
above a mass of 125 GeV/$c^2$, there should be no problem in measuring the
predicted $\sim 50\%$ asymmetry to about 10\% of its value. Thus, effects of
new physics should be at least this large to be observable.

An order-of-magnitude estimate is possible on the basis of the amplitude
(\ref{eqn:amp}) and the couplings (\ref{eqn:coup}).  The extra factor of $x
\equiv \sin^2 \theta_W$ in $\gth^2$ relative to $g_Z^2$ occurs for any $Z'$
coupling to a U(1) charge.  For a lepton pair mass of 125 GeV/$c^2$, the
fractional effect on the scattering amplitude is no larger than $x(125~
\G/c^2/M_{Z'})^2$, or a couple of percent for $M_{Z'} = 400$ GeV/$c^2$.  In the
next section we shall illustrate this estimate with some explicit examples.
\bigskip

\centerline{\bf III.   EFFECTS OF A 500 GeV $Z'$}
\bigskip

We assume $M_{Z'} = 500~\G/c^2$, beyond the published mass limits
\cite{CDFZlims}, and calculate the expected forward-backward asymmetries in $f
\bar f \to e^- e^+ + \ldots$ for $f = u,~d,~\mu$.  (The most convenient
formalism for incorporating parton-level results into a calculation with
up-to-date structure functions and realistic detector acceptance is given in
Ref.~\cite{offpk}.  The case $f = \mu$ is relevant for the process $e^- e^+ \to
\mu^- \mu^+$ at a next-generation linear collider.)  The results are shown in
Figs.~1--3.

\begin{figure}
\centerline{\epsfysize = 5in \epsffile{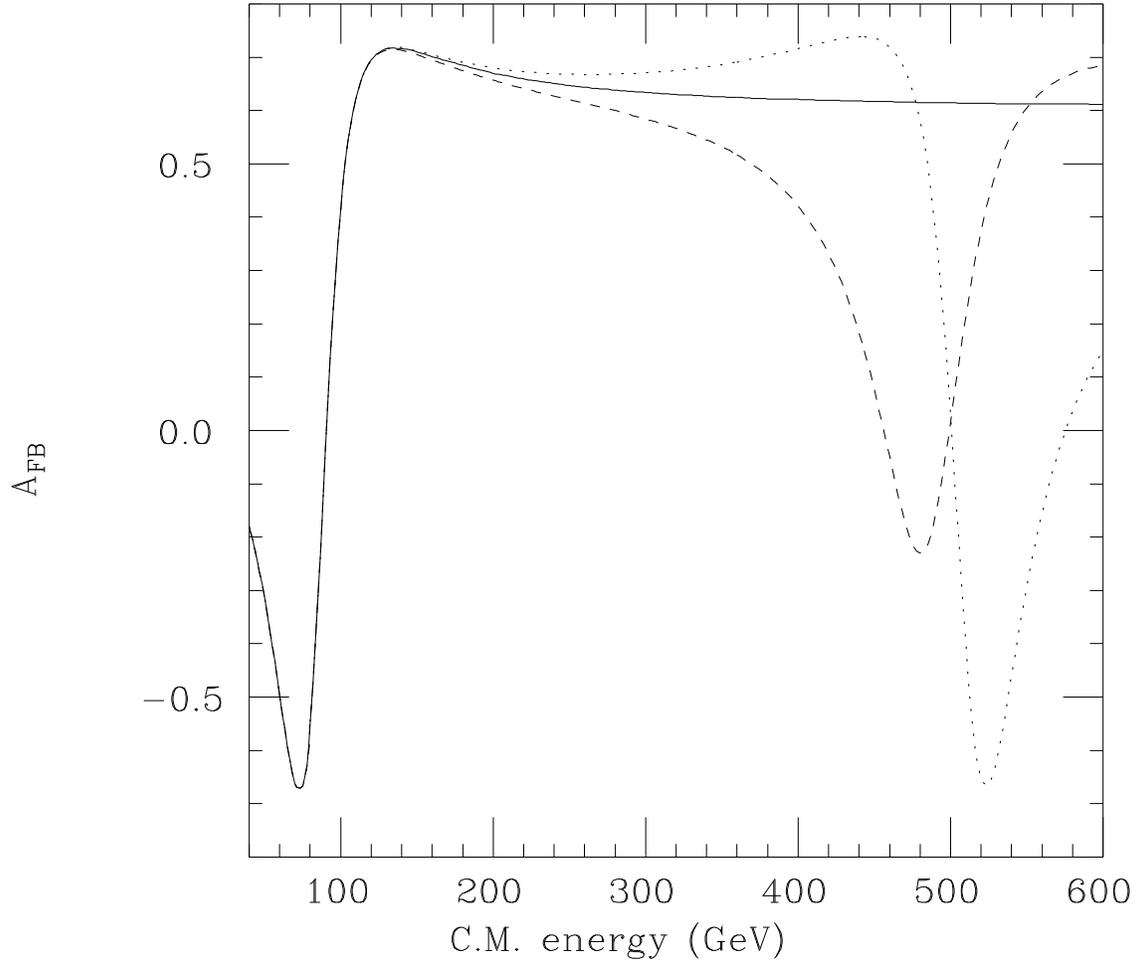}}
\caption{Parton-level forward-backward asymmetries for $u \bar u \to e^-
e^+$.  Solid line:  standard model.  Dashed line: 500 GeV/$c^2~Z_\chi$ added.
Dotted line:  500 GeV/$c^2~Z_\psi$ added.  A $Z_I$ does not couple to $u$
quarks and does not change the standard model prediction.}
\end{figure}

In these figures the first point to note is the relative insensitivity to a 500
GeV $Z'$ of physics at or below 200 GeV.  This is in part a feature of our
assumption that the standard $Z$ and the $Z'$ are very weakly mixed with one
another, and in part stems from the relatively weak coupling assumed for the
$Z'$.

When the lepton pair mass approaches $M_{Z'}$, the characteristic interferences
differ substantially from one another for various kinds of $Z'$.  These
patterns can be very helpful in diagnosing the nature of a new neutral gauge
boson \cite{offpk}.

The asymmetries in $u \bar u \to e^- e^+$ are the same for the standard model
and when a $Z_I$ is added, since that boson does not couple to $u$ quarks.

The asymmetries when a $Z_\psi$ is added are very small at the pole, since a
$Z\psi$ couples purely axially to the ordinary quarks and leptons.

\begin{figure}
\centerline{\epsfysize = 5in \epsffile{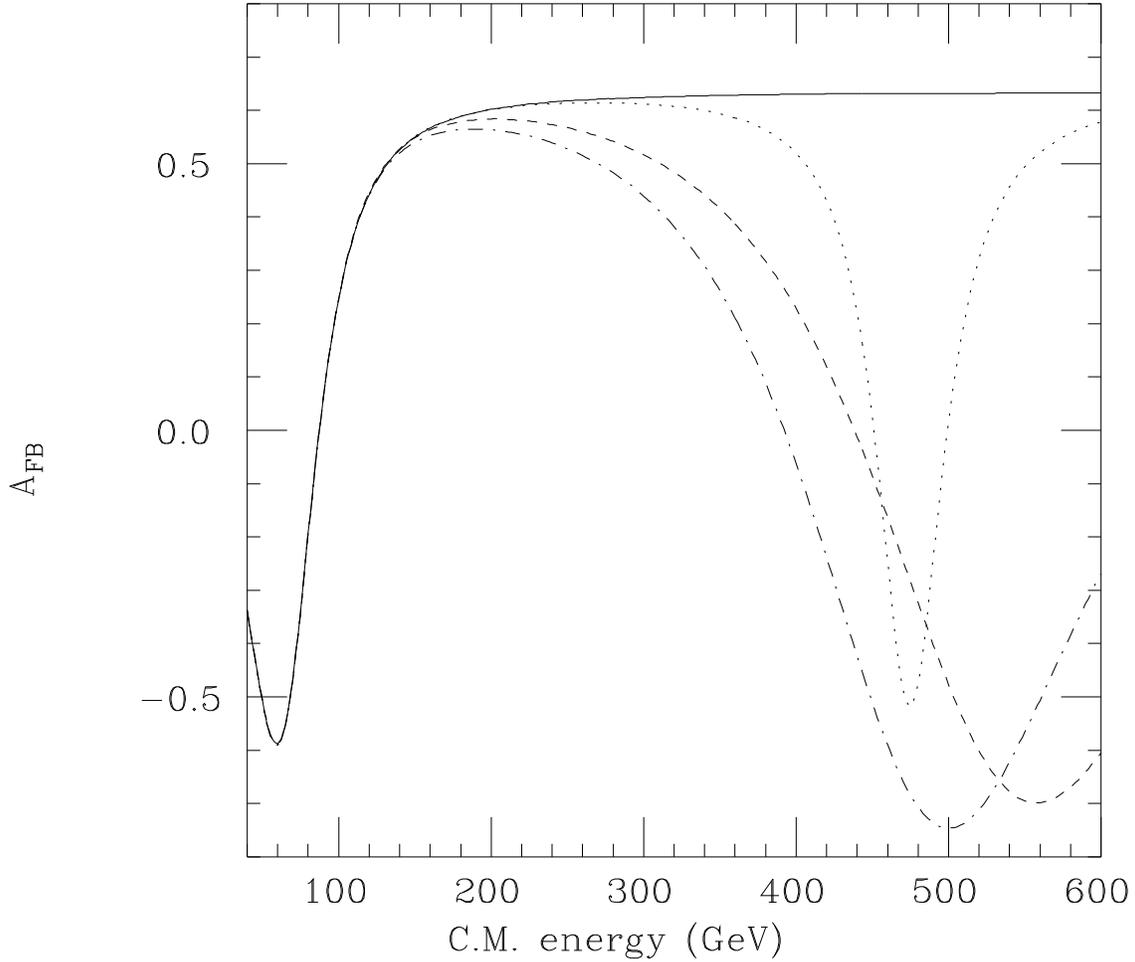}}
\caption{Parton-level forward-backward asymmetries for $d \bar d \to e^-
e^+$.  Solid line:  standard model.  Dashed line: 500 GeV/$c^2~Z_\chi$ added.
Dotted line:  500 GeV/$c^2~Z_\psi$ added.  Dot-dashed line: 500 GeV/$c^2~
Z_I$ added.}
\end{figure}

The most likely place where asymmetries due to a $Z'$ will first be observed is
at the pole mass.  Accordingly, in Fig.~4 we have shown the asymmetries for $f
\bar f \to e^- e^+$ at a subenergy of 500 GeV due to a $Z'$ of mass 500
GeV/$c^2$, parametrized by the angle $\theta$ in Eq.~(\ref{eqn:th}).  Also
shown is the ratio
\beq
r \equiv \frac{(C_L^u)^2 + (C_R^u)^2}{(C_L^d)^2 + (C_R^d)^2}
\eeq
describing the relative strengths of $u$ and $d$ quark couplings.  The ratio
$r$ reaches its maximum of 2 at $\theta = {\rm arcsin}(\sqrt{5/2}/4) \simeq
23.28^{\circ}$, and vanishes for $\theta = \theta_I \equiv {\rm
arctan}(-\sqrt{5/3}) \simeq 127.78^{\circ}$, i.e., for $Z' = Z_I$.  In the
latter case, $Z'$ production in hadronic colliders is due entirely to $d \bar
d$ annihilation, and the large negative asymmetry in $d \bar d \to e^- e^+$
will be reflected in a similar asymmetry in $p \bar p \to e^- e^+ + \ldots$. To
some extent this is also true of the $Z_\chi$, for which $r$ is only 1/5. With
no more than about a dozen events of $p \bar p \to Z_{\chi,I} \to \ell^-
\ell^+$, it should therefore be possible to demonstrate a convincing
forward-backward asymmetry very different from that due to the photon and $Z$.

\begin{figure}
\centerline{\epsfysize = 5in \epsffile{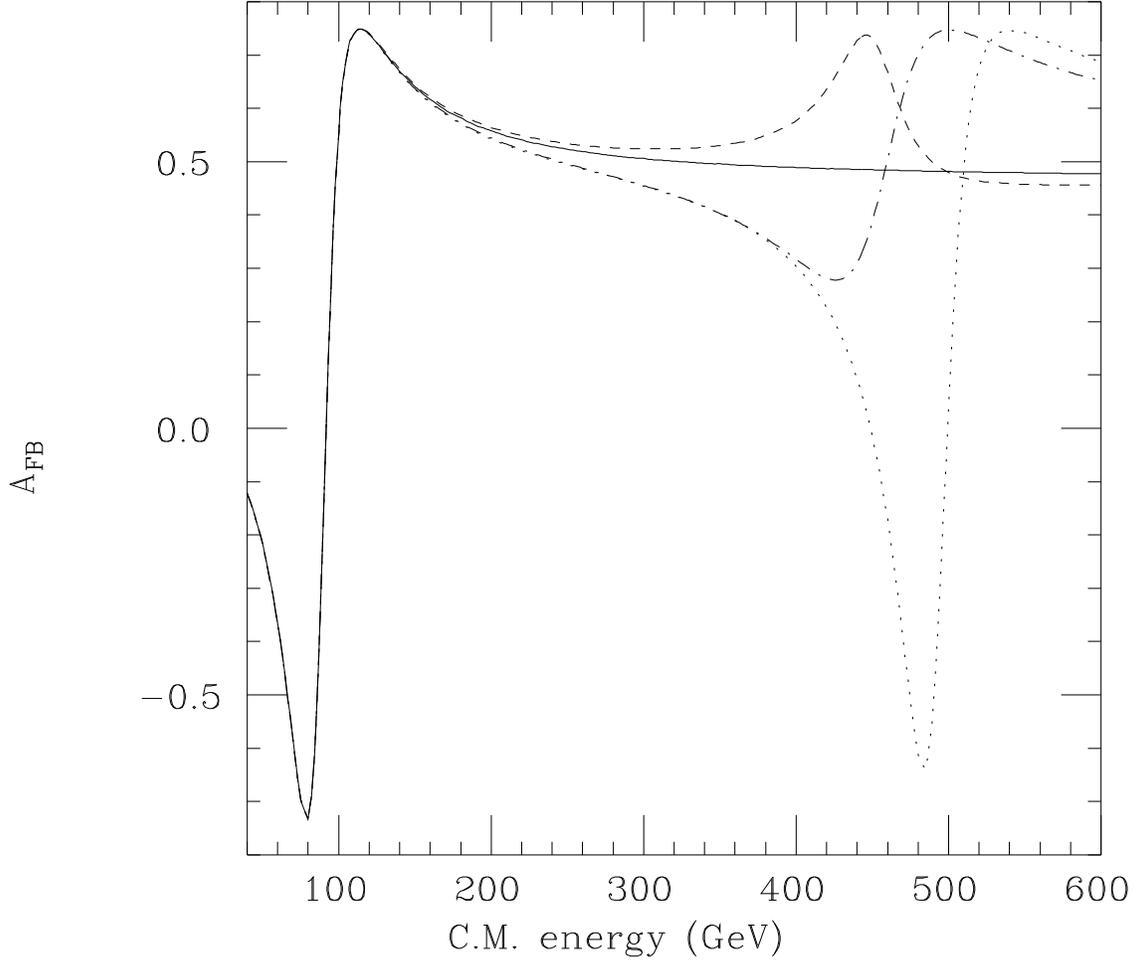}}
\caption{Same as previous figure for $\mu^- \mu^+ \to e^- e^+$.}
\end{figure}

\begin{figure}
\centerline{\epsfysize = 7in \epsffile{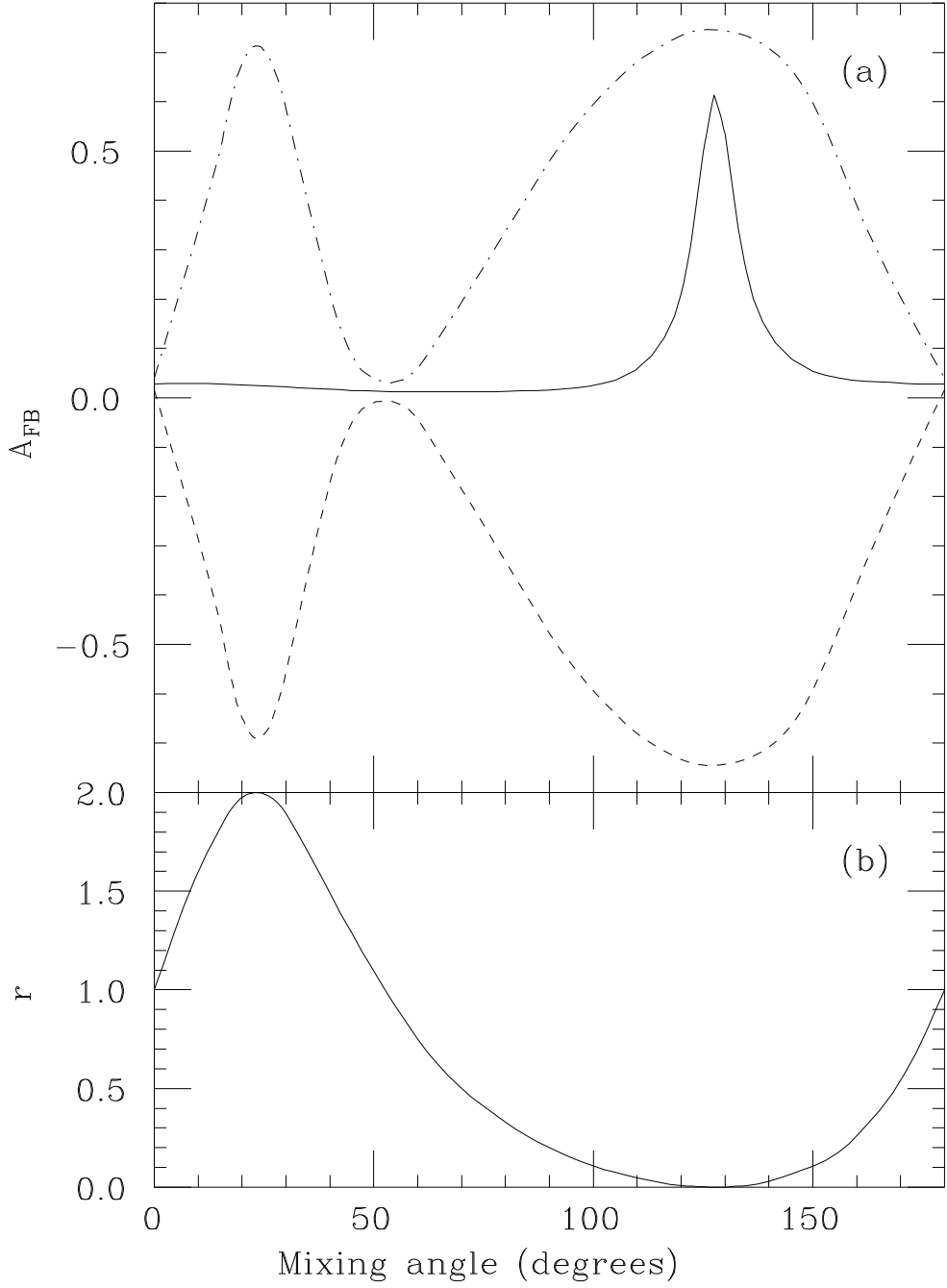}}
\caption{ Parameters as a function of the mixing angle $\theta$ for a 500
GeV/$c^2$ $Z'$, as measured on the peak.  (a) Forward-backward asymmetries at
the parton level.  Solid line:  $u \bar u \to e^- e^+$; dashed line: $d \bar d
\to e^- e^+$; dotdashed line: $\mu^- \mu^+ \to e^- e^+$.  (b)  Ratio $r$ of
sums of squares of $u$ and $d$ production coefficients.}
\end{figure}

The suppression of the couplings of a $Z'$ to $u$ quarks over a wide range of
$\theta$ indicated in Fig.~4(b) implies that the {\it decays} of the $Z'$ may
be enriched in the charge $-1/3$ quarks $d,~s,~b$.  Moreover, as one can see
from the couplings in Table I, there is the potential for $|C_L^d| \ll
|C_R^d|$, leading to a characteristic forward-peaking in the process $d \bar d
\to Z' \to (d,~s,~{\rm or}~b) + (\bar d,~ \bar s,~{\rm or}~\bar b)$. It may be
worth while looking for such an effect by studying the angular distribution of
jets produced at high transverse momentum using momentum-weighted jet charge
methods.

The asymmetries at the pole for $d \bar d \to e^- e^+$ and $\mu^- \mu^+ \to e^-
e^+$ vanish at $\theta = 0$ (where the couplings to the $Z'$ are purely axial)
and $\theta = {\rm arctan}(\sqrt{5/3}) \simeq 52.24^{\circ}$, where they are
purely vector.  Aside from small effects due to interference with the photon
and $Z$, these asymmetries are always of the same sign:  negative for $d \bar d
\to e^- e^+$ and positive for $\mu^- \mu^+ \to e^- e^+$.

The peak in the asymmetry for $u \bar u \to e^- e^+$ at $\theta = \theta_I$ is
due to the vanishing of the $Z'$ contribution, whereupon the large positive
standard model value is attained.

\bigskip

\centerline{\bf IV.  REMARKS ON COMPOSITENESS}
\bigskip

Analyses similar to those carried out for extra neutral gauge bosons can be
performed when the $f \bar f \to e^- e^+$ scattering amplitude is modified by a
term due to exchange of a very heavy strongly-coupled boson, such as that which
would arise in certain composite models.  Specifically, in Ref.~\cite{RS} the
amplitude for $f_L \bar f \to e_L^- e^+$ was taken to be
$$
A_{LL} \equiv A(f_L \bar f \to e^-_L e^+) = -Q_f e^2 + \frac{\shat}
{\shat - m_Z^2 + i M_Z \Gamma_Z} C_L^Z(f) C_L^Z(e)
$$
\beq
+ \xi \frac{g_Y^2}{2}\frac{\shat}{\shat-M_V^2}~~~,
\eeq
where $\xi = \pm 1$, and $g_Y$ is a new coupling constant associated with the
exchange of a hypothetical vector meson $V$ with mass taken to equal 2
TeV/$c^2$.  Other scattering amplitudes were taken to have standard form. The
fractional effect on the amplitude should be of order $(\alpha_Y/2
\alpha)(\shat/M_V^2)$, or roughly $\alpha_Y/4$ for $\sqrt{\shat} = 125$ GeV.
Thus with a 10\% measurement of asymmetries above a lepton pair mass of 125
GeV/$c^2$ one could exclude a value of $\alpha_Y (2~{\rm TeV}/M_V)^2$ above
0.4, or a value of $M_V$ below about 3.2 TeV/$c^2$ for $\alpha_Y = 1$. This is
comparable to values set in other searches at present.

In contrast to models with extra neutral gauge bosons, the composite model
discussed in Ref.~\cite{RS} also predicts appreciable effects in
charged-current lepton pair production, such as $p \bar p \to e^- + \bar \nu +
\ldots$.  Even with $\alpha_Y(2~{\rm TeV}/M_V)^2$ as small as 0.1, one expects
deviations by up to a factor of two from the standard model predictions for the
cross section for production of single charged leptons above a transverse
momentum of 200 GeV/$c$.
\newpage

\centerline{\bf V.  CONCLUSIONS}
\bigskip

We have considered the effects of two new types of physics on forward-backward
asymmetries of high-mass charged lepton pairs produced in proton-antiproton
collisions at the Fermilab Tevatron.

(1) A new neutral gauge boson (``$Z'$'') can modify scattering amplitudes at
the parton level, but its effects will probably not be visible through
interference with the photon and $Z$ contributions in the current sample of
data.  A handful of events at the peak of a new $Z$ can, however, already
exhibit asymmetries strikingly different from those in the standard model for a
range of possible forms of the $Z'$.

(2) Composite models can give rise to effective contact terms, whose fractional
contribution to electroweak amplitudes can probably be ruled out or discovered
at the 10\% level in the current round of experiments.  For coupling strengths
of order $\alpha_Y = 1$, these would correspond to compositeness scales of
order 5 TeV, comparable to those probed in other current experiments.  In one
class of models \cite{RS}, such terms are expected to be particularly prominent
in {\it charged-current} lepton pair production, such as is studied in searches
for excited $W$ bosons \cite{CDFWlims}.
\bigskip

\centerline{\bf ACKNOWLEDGMENTS}
\bigskip

I am grateful to H. Frisch for stimulating me to undertake this investigation
and for helpful discussions. This work was supported in part by the United
States Department of Energy under Grants No. DE AC02 76CH03000 and DE FG02
90ER40560.
\bigskip

\def \ajp#1#2#3{Am. J. Phys. {\bf#1}, #2 (#3)}
\def \apny#1#2#3{Ann. Phys. (N.Y.) {\bf#1}, #2 (#3)}
\def \app#1#2#3{Acta Phys. Polonica {\bf#1}, #2 (#3)}
\def \arnps#1#2#3{Ann. Rev. Nucl. Part. Sci. {\bf#1}, #2 (#3)}
\def \cmts#1#2#3{Comments on Nucl. Part. Phys. {\bf#1}, #2 (#3)}
\def \cn{Collaboration}
\def \cp89{{\it CP Violation,} edited by C. Jarlskog (World Scientific,
Singapore, 1989)}
\def \efi{Enrico Fermi Institute Report No. EFI}
\def \f79{{\it Proceedings of the 1979 International Symposium on Lepton and
Photon Interactions at High Energies,} Fermilab, August 23-29, 1979, ed. by
T. B. W. Kirk and H. D. I. Abarbanel (Fermi National Accelerator Laboratory,
Batavia, IL, 1979}
\def \hb87{{\it Proceeding of the 1987 International Symposium on Lepton and
Photon Interactions at High Energies,} Hamburg, 1987, ed. by W. Bartel
and R. R\"uckl (Nucl. Phys. B, Proc. Suppl., vol. 3) (North-Holland,
Amsterdam, 1988)}
\def \ib{{\it ibid.}~}
\def \ibj#1#2#3{~{\bf#1}, #2 (#3)}
\def \ichep72{{\it Proceedings of the XVI International Conference on High
Energy Physics}, Chicago and Batavia, Illinois, Sept. 6 -- 13, 1972,
edited by J. D. Jackson, A. Roberts, and R. Donaldson (Fermilab, Batavia,
IL, 1972)}
\def \ijmpa#1#2#3{Int. J. Mod. Phys. A {\bf#1}, #2 (#3)}
\def \ite{{\it et al.}}
\def \jpb#1#2#3{J.~Phys.~B~{\bf#1}, #2 (#3)}
\def \lkl87{{\it Selected Topics in Electroweak Interactions} (Proceedings of
the Second Lake Louise Institute on New Frontiers in Particle Physics, 15 --
21 February, 1987), edited by J. M. Cameron \ite~(World Scientific, Singapore,
1987)}
\def \ky85{{\it Proceedings of the International Symposium on Lepton and
Photon Interactions at High Energy,} Kyoto, Aug.~19-24, 1985, edited by M.
Konuma and K. Takahashi (Kyoto Univ., Kyoto, 1985)}
\def \mpla#1#2#3{Mod. Phys. Lett. A {\bf#1}, #2 (#3)}
\def \nc#1#2#3{Nuovo Cim. {\bf#1}, #2 (#3)}
\def \np#1#2#3{Nucl. Phys. {\bf#1}, #2 (#3)}
\def \PDG{Particle Data Group, L. Montanet \ite, \prd{50}{1174}{1994}}
\def \pisma#1#2#3#4{Pis'ma Zh. Eksp. Teor. Fiz. {\bf#1}, #2 (#3) [JETP Lett.
{\bf#1}, #4 (#3)]}
\def \pl#1#2#3{Phys. Lett. {\bf#1}, #2 (#3)}
\def \pla#1#2#3{Phys. Lett. A {\bf#1}, #2 (#3)}
\def \plb#1#2#3{Phys. Lett. B {\bf#1}, #2 (#3)}
\def \pr#1#2#3{Phys. Rev. {\bf#1}, #2 (#3)}
\def \prc#1#2#3{Phys. Rev. C {\bf#1}, #2 (#3)}
\def \prd#1#2#3{Phys. Rev. D {\bf#1}, #2 (#3)}
\def \prl#1#2#3{Phys. Rev. Lett. {\bf#1}, #2 (#3)}
\def \prp#1#2#3{Phys. Rep. {\bf#1}, #2 (#3)}
\def \ptp#1#2#3{Prog. Theor. Phys. {\bf#1}, #2 (#3)}
\def \rmp#1#2#3{Rev. Mod. Phys. {\bf#1}, #2 (#3)}
\def \rp#1{~~~~~\ldots\ldots{\rm rp~}{#1}~~~~~}
\def \si90{25th International Conference on High Energy Physics, Singapore,
Aug. 2-8, 1990}
\def \slc87{{\it Proceedings of the Salt Lake City Meeting} (Division of
Particles and Fields, American Physical Society, Salt Lake City, Utah, 1987),
ed. by C. DeTar and J. S. Ball (World Scientific, Singapore, 1987)}
\def \slac89{{\it Proceedings of the XIVth International Symposium on
Lepton and Photon Interactions,} Stanford, California, 1989, edited by M.
Riordan (World Scientific, Singapore, 1990)}
\def \smass82{{\it Proceedings of the 1982 DPF Summer Study on Elementary
Particle Physics and Future Facilities}, Snowmass, Colorado, edited by R.
Donaldson, R. Gustafson, and F. Paige (World Scientific, Singapore, 1982)}
\def \smass90{{\it Research Directions for the Decade} (Proceedings of the
1990 Summer Study on High Energy Physics, June 25--July 13, Snowmass,
Colorado),
edited by E. L. Berger (World Scientific, Singapore, 1992)}
\def \tasi90{{\it Testing the Standard Model} (Proceedings of the 1990
Theoretical Advanced Study Institute in Elementary Particle Physics, Boulder,
Colorado, 3--27 June, 1990), edited by M. Cveti\v{c} and P. Langacker
(World Scientific, Singapore, 1991)}
\def \yaf#1#2#3#4{Yad. Fiz. {\bf#1}, #2 (#3) [Sov. J. Nucl. Phys. {\bf #1},
#4 (#3)]}
\def \zhetf#1#2#3#4#5#6{Zh. Eksp. Teor. Fiz. {\bf #1}, #2 (#3) [Sov. Phys. -
JETP {\bf #4}, #5 (#6)]}
\def \zpc#1#2#3{Zeit. Phys. C {\bf#1}, #2 (#3)}
\def \zpd#1#2#3{Zeit. Phys. D {\bf#1}, #2 (#3)}

\end{document}